\documentclass[aps,pra,noshowpacs,twocolumn,reprint,superscriptaddress,floatfix]{revtex4-1}

\pdfoutput=1
\usepackage{xcolor, soul}
\usepackage{amsmath,amssymb}
\usepackage{multirow}
\usepackage{hyperref,graphicx}

\newcommand{\ket}[1]{\left| #1 \right>} 
\newcommand{\braket}[2]{\left< #1 \vphantom{#2} \right| \left. #2 \vphantom{#1} \right>}

\begin{document}
\title{Photonic Newton's Cradle for Remote Energy Transport}

\author{Zhen Feng}
\thanks{These authors contributed equally.}
\affiliation{State Key Laboratory of Advanced Optical Communication Systems and Networks, School of Physics and Astronomy, Shanghai Jiao Tong University, Shanghai 200240, China}
\affiliation{Synergetic Innovation Center of Quantum Information and Quantum Physics, University of Science and Technology of China, Hefei, Anhui 230026, China}

\author{Zhen-Wei Gao}
\thanks{These authors contributed equally.}
\affiliation{State Key Laboratory of Advanced Optical Communication Systems and Networks, School of Physics and Astronomy, Shanghai Jiao Tong University, Shanghai 200240, China}
\affiliation{Synergetic Innovation Center of Quantum Information and Quantum Physics, University of Science and Technology of China, Hefei, Anhui 230026, China}

\author{Lian-Ao Wu}
\affiliation{Ikerbasque, Basque Foundation for Science, 48011 Bilbao, and Department of Theoretical Physics and History of Science, University of Basque Country, Bilbao 48080, Spain}

\author{Hao Tang}
\affiliation{State Key Laboratory of Advanced Optical Communication Systems and Networks, School of Physics and Astronomy, Shanghai Jiao Tong University, Shanghai 200240, China}
\affiliation{Synergetic Innovation Center of Quantum Information and Quantum Physics, University of Science and Technology of China, Hefei, Anhui 230026, China}

\author{Ke Sun}
\affiliation{State Key Laboratory of Advanced Optical Communication Systems and Networks, School of Physics and Astronomy, Shanghai Jiao Tong University, Shanghai 200240, China}

\author{Cheng-Qiu Hu}
\affiliation{State Key Laboratory of Advanced Optical Communication Systems and Networks, School of Physics and Astronomy, Shanghai Jiao Tong University, Shanghai 200240, China}
\affiliation{Synergetic Innovation Center of Quantum Information and Quantum Physics, University of Science and Technology of China, Hefei, Anhui 230026, China}

\author{Yao Wang}
\affiliation{State Key Laboratory of Advanced Optical Communication Systems and Networks, School of Physics and Astronomy, Shanghai Jiao Tong University, Shanghai 200240, China}
\affiliation{Synergetic Innovation Center of Quantum Information and Quantum Physics, University of Science and Technology of China, Hefei, Anhui 230026, China}

\author{Zhan-Ming Li}
\affiliation{State Key Laboratory of Advanced Optical Communication Systems and Networks, School of Physics and Astronomy, Shanghai Jiao Tong University, Shanghai 200240, China}
\affiliation{Synergetic Innovation Center of Quantum Information and Quantum Physics, University of Science and Technology of China, Hefei, Anhui 230026, China}

\author{Xiao-Wei Wang}
\affiliation{State Key Laboratory of Advanced Optical Communication Systems and Networks, School of Physics and Astronomy, Shanghai Jiao Tong University, Shanghai 200240, China}

\author{Yuan Chen}
\affiliation{State Key Laboratory of Advanced Optical Communication Systems and Networks, School of Physics and Astronomy, Shanghai Jiao Tong University, Shanghai 200240, China}
\affiliation{Synergetic Innovation Center of Quantum Information and Quantum Physics, University of Science and Technology of China, Hefei, Anhui 230026, China}

\author{En-Ze Zhang}
\affiliation{State Key Laboratory of Advanced Optical Communication Systems and Networks, School of Physics and Astronomy, Shanghai Jiao Tong University, Shanghai 200240, China}

\author{Zhi-Qiang Jiao}
\affiliation{State Key Laboratory of Advanced Optical Communication Systems and Networks, School of Physics and Astronomy, Shanghai Jiao Tong University, Shanghai 200240, China}
\affiliation{Synergetic Innovation Center of Quantum Information and Quantum Physics, University of Science and Technology of China, Hefei, Anhui 230026, China}

\author{Xiao-Yun Xu}
\affiliation{State Key Laboratory of Advanced Optical Communication Systems and Networks, School of Physics and Astronomy, Shanghai Jiao Tong University, Shanghai 200240, China}
\affiliation{Synergetic Innovation Center of Quantum Information and Quantum Physics, University of Science and Technology of China, Hefei, Anhui 230026, China}

\author{Jun Gao}
\affiliation{State Key Laboratory of Advanced Optical Communication Systems and Networks, School of Physics and Astronomy, Shanghai Jiao Tong University, Shanghai 200240, China}
\affiliation{Synergetic Innovation Center of Quantum Information and Quantum Physics, University of Science and Technology of China, Hefei, Anhui 230026, China}

\author{Ai-Lin Yang}
\affiliation{State Key Laboratory of Advanced Optical Communication Systems and Networks, School of Physics and Astronomy, Shanghai Jiao Tong University, Shanghai 200240, China}
\affiliation{Synergetic Innovation Center of Quantum Information and Quantum Physics, University of Science and Technology of China, Hefei, Anhui 230026, China}

\author{Xian-Min Jin}
\thanks{xianmin.jin@sjtu.edu.cn}
\affiliation{State Key Laboratory of Advanced Optical Communication Systems and Networks, School of Physics and Astronomy, Shanghai Jiao Tong University, Shanghai 200240, China}
\affiliation{Synergetic Innovation Center of Quantum Information and Quantum Physics, University of Science and Technology of China, Hefei, Anhui 230026, China}

\maketitle

\textbf{Energy transport is of central importance in understanding a wide variety of transitions of physical states in nature. Recently, the coherence and noise have been identified for their existence and key roles in energy transport processes, for instance, in a photosynthesis complex \cite{Engel2007,Mohseni2018}, DNA \cite{Rieper2010}, and odor sensing \cite{Arndt2009} etc, of which one may have to reveal the inner mechanics in the quantum regime. Here we present an analog of Newton's cradle by manipulating a boundary-controlled chain on a photonic chip. Long-range interactions can be mediated by a long chain composed of $21$ strongly coupled sites, where single-photon excitations are transferred between two remote sites via simultaneous control of inter-site weak and strong couplings. We observe a high retrieval efficiency in both uniform and defect-doped chain structures. Our results may offer a flexible approach to Hamiltonian engineering beyond geometric limitation, enabling the design and construction of quantum simulators on demand.}
\vskip -3.5mm
\section{INTRODUCTION}
\vskip -3.5mm
\noindent Energy transport, unveiling the evolution of physical states and the nature of particle interactions has long been discussed. A quantum particle that can also be represented as a wave function possesses quantum-inherent interference and has a superposition of many locations or paths, and quantum-involved energy transport in the quantum regime is stepping into fascinating frontiers\cite{Engel2007,Arndt2009,Mohseni2018,Rieper2010,Bentivegna2015,Pitsios2017}.

A ballistic spread has been performed when a quantum particle propagates in ideally ordered lattices~\cite{Kempe2003}. However,  the energy mismatch between sites in practical system Hamiltonian may lead to localization~\cite{Schwartz2007,Lahini2008} and the interaction with environment may result in decoherence~\cite{Maziero2009}. For instance, unexpected accumulated phases or amplitudes may cause the suppression of broadening wave package \cite{Lahini2008}.

A chain, linearly arranged sites between nearest coupled neighbors, has been considered for energy transport between information terminals at a distance, which is crucial in constructing large-scale quantum information networks. Two types of chains, fully-engineered chain and boundary-controlled chain, have been proposed to realize energy transport. The fully-engineered chain requires an accurate coupling modulation of $\sqrt{i(N-i)}$ on the structural configuration and a perfect spatial mirror image of the injection after a half period to perform a perfect retrieval of photons \cite{Christandl2004,Bellec2012,Chapman2016}. The transfer efficiency is highly dependent on the manufacturing precision associated with both external and internal regularity. The high parameter sensitivity demanded in theory and unavoidable imperfections in practice make the chain hard to scale up.

In contrast, a boundary-controlled chain as a new model was proposed and it required a minimal control while possessing a well-behaved robustness \cite{Wojcik2005}. A series of linearly arranged and strongly coupled sites form a chain to bridge two remote sites, denoted as sending and receiving sites. The two sites are coupled to the end of the chain with a very weak coupling \cite{Yao2011,Wang2013}. A quantum particle injected into the sending site can propagate through the boundary-controlled chain and go back and forth to the receiving site, with a fashion of energy transport like Newton's cradle. A straightforward model (with only two parameters of strong and weak couplings) and its intrinsic robustness makes the protocol of boundary-controlled chain a promising candidate for energy transport. The boundary-controlled chain models were widely investigated and have been applied to different physical systems \cite{Wu2002,Plenio2005,Shi2005,Li2005,CamposVenuti2007,Oh2012,Stolze2014,Stolze2016,Vieira2018}, which, however, have not been experimentally demonstrated yet.

\begin{figure*}[htb!!]
\includegraphics[width=0.98\textwidth]{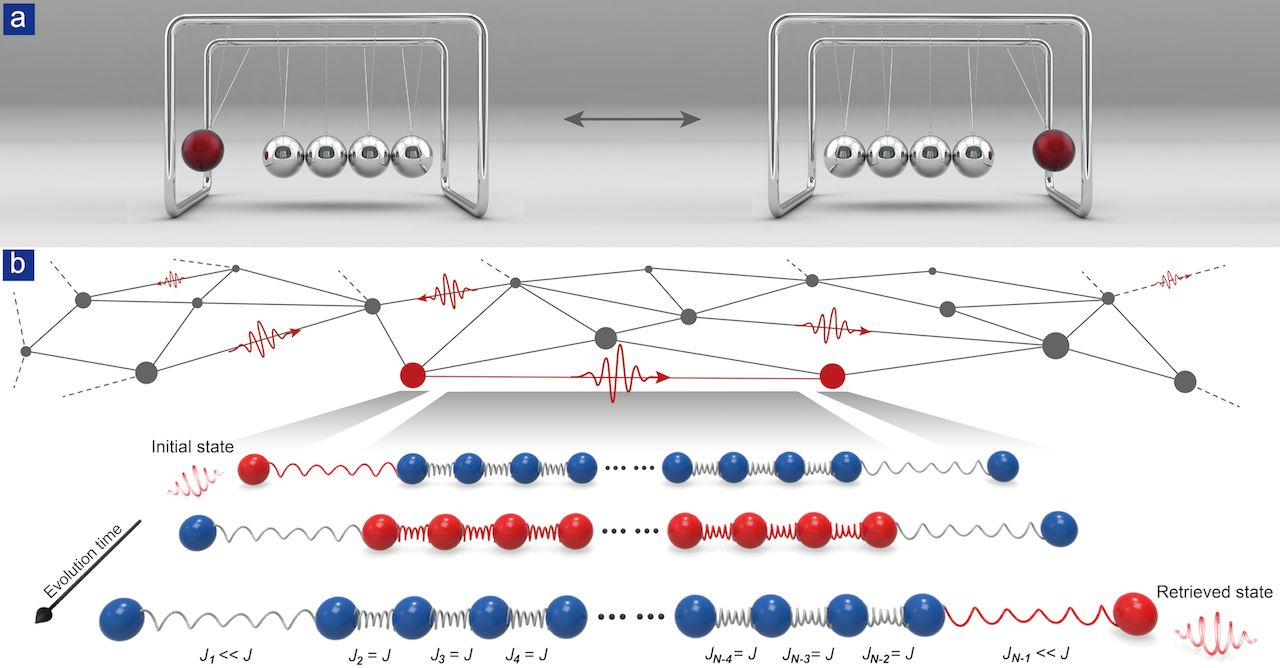}
\caption{\textbf{Energy transport in Newton's cradle and its analog in photonic system.} \textbf{a.} Energy exchange and interaction mechanics in Newton's cradle. The energy can be transferred from the leftmost ball to the rightmost one mediated by a chain of balls, in effect, which is equivalent to the scenario that the two balls interact each other directly. \textbf{b.} Energy transport between remote sites in a quantum network. The photonic analog of Newton's cradle is realized by a boundary-controlled chain. The chain consists of weak coupling $\alpha J$ in two ends and a chain of strongly coupling $J$ in the middle, which gives an equivalent coupling for the two remote sites. The obtained long-range interaction could enable an faithful transfer of quantum states.  an initial state can be transferred to the remote receiving site   faithfully and therefore can serve like a photonic Newton's cradle for enabling interactions between remote sites in a quantum network.}
\label{fig:figure1}
\end{figure*}
\vskip -3.5mm
\vskip -3.5mm
\section{RESULTS}
\vskip -3.5mm
\subsection{Photonic analog of Newton's cradle}
\vskip -3.5mm
In this paper, we prototype waveguides as the sites of the chain by using femtosecond laser direct writing technique \cite{Szameit2007,Feng2016,Tang2018,Tang2018a} and successfully map the boundary-controlled chain onto a photonic chip (see Appendix A). As shown in Fig. \ref{fig:figure1}, with the time evolution, a photon is launched on the left sending site, weakly coupled to a head-tail strongly-coupled chain, then also weakly coupled to the right receiving site. Such a boundary-controlled chain can transfer single photons faithfully and therefore can serve like a photonic Newton's cradle for enabling interactions between remote sites in a quantum network.

The Hamiltonian is composed of strong couplings $J_n=J, (n=2,3,\dots, N-2)$ and weak couplings $J_1=J_{N-1}=\alpha J, (\alpha \ll 1)$. The parameter $\alpha$ reflects the differential strength between strong and weak couplings. A single photon state $\ket{\Psi_0}$ is injected in the sending site. Initial state of the N-body system which can be presented as $\ket{\Phi(t=0)}=\ket{\Psi_0}\otimes\ket{00\dots0}$ evolves with the Hamiltonian:
\begin{equation}
H=J_{1} a_{1}^{\dagger} a_{2} + J_{N-1} a_{N-1}^{\dagger} a_{N} +  \sum_{i=2}^{N-2}J a_{i}^{\dagger} a_{i+1} + c.c.
\label{eq:Equation1}
\end{equation}
Here we employ a coupled mode approach to experimentally characterizing the coupling strength between two waveguides \cite{Szameit2007}.

The state of the system at a given time $t$ is $\ket{\Phi(t)}=e^{-iHt/\hbar}\ket{\Phi(t=0)}$. Given the parameters $N$ and $\alpha$, we have the highest efficiency $\eta=1-O(\alpha^2 N)$ in the optimal receiving time 
\begin{equation}
\tau =\frac{\pi\sqrt{N-2}}{2(\alpha J)}
\label{time}
\end{equation}
where the capital $O$ notation is made use of keeping track of the leading term that dominates scaling \cite{Nielsen2002}. In fact, the optimal receiving time $\tau$ and transfer efficiency $\eta$ are a trade-off. A high efficiency is at the cost of a weak coupling( $\alpha <1/\sqrt{N}$) and a faithful transfer can be obtained asymptotically with $\alpha$ nearly vanishing~\cite{Zwick2012}.

\begin{figure*}[htb!]
\includegraphics[width=0.8\textwidth]{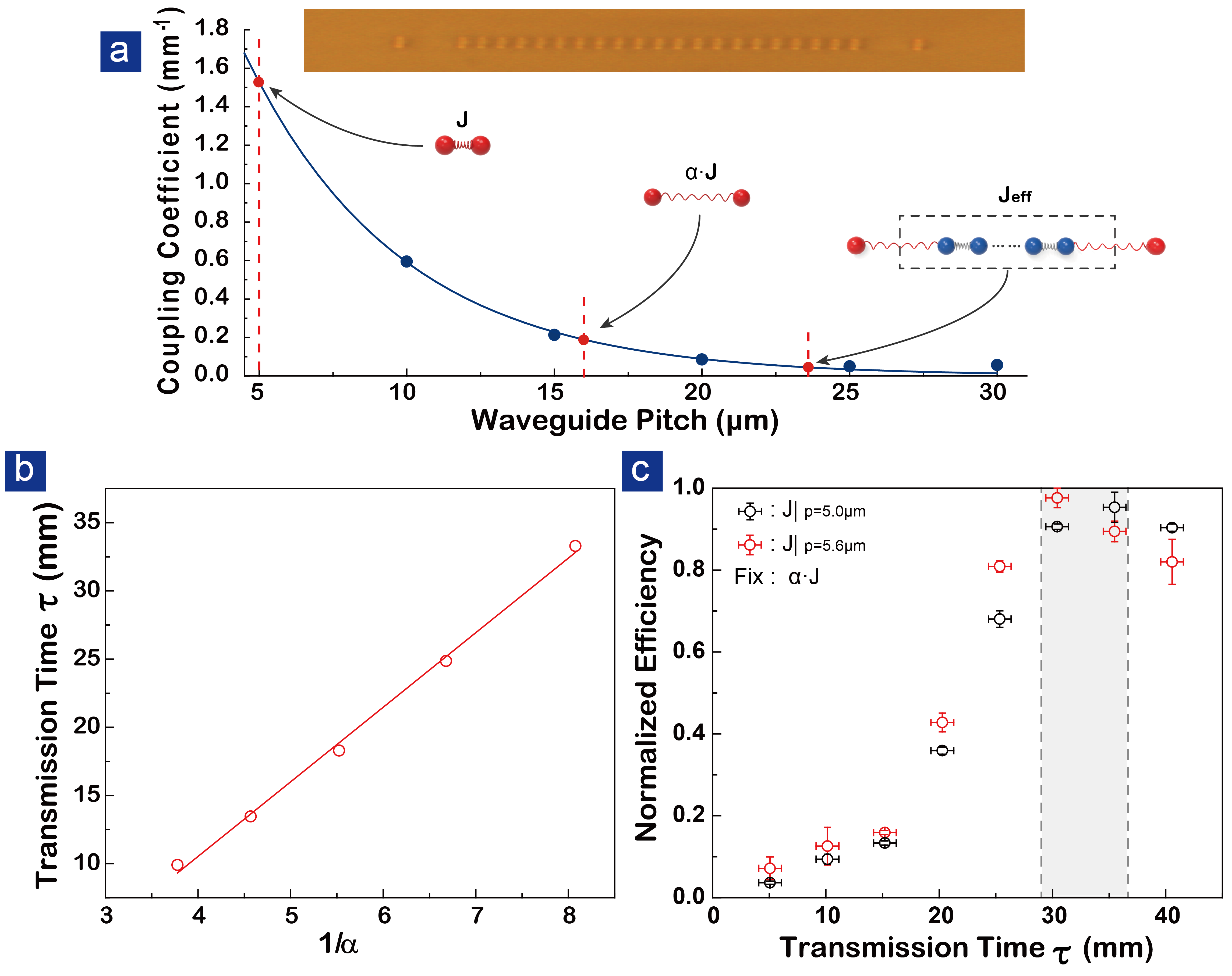}
\caption{\textbf{Design and implementation of the boundary-controlled chain.} \textbf{a.} Characterized coupling coefficients and their exponential dependence on the waveguide pitches. An effective coupling coefficient can be designed by choosing appropriate $\alpha J$ and $J$ respectively. The blue dots show experimental measurements. The red dashed lines and spots mark the parameters mainly adopted in this work. The inset is the cross section of a set of waveguide array with designed pitches. \textbf{b.} The linear relation, as indicated in Eq. \ref{time}, is observed by scanning $1/\alpha$ while keeping the product $\alpha J$ constant. Here, the transfer time $\tau$, in unit of $mm$, is defined by the evolution length of photons in the waveguide array. \textbf{c.} The robustness of the model is demonstrated by showing the insensitivity of the transfer time on the strong coupling $J$. The small shift can be attributed to the unavoidable experimental imperfections and the fact that the condition $\alpha \ll 1$ is not strictly satisfied.}
\label{fig:figure2}
\end{figure*}

\begin{figure*}[htb!]
\includegraphics[width=0.75\textwidth]{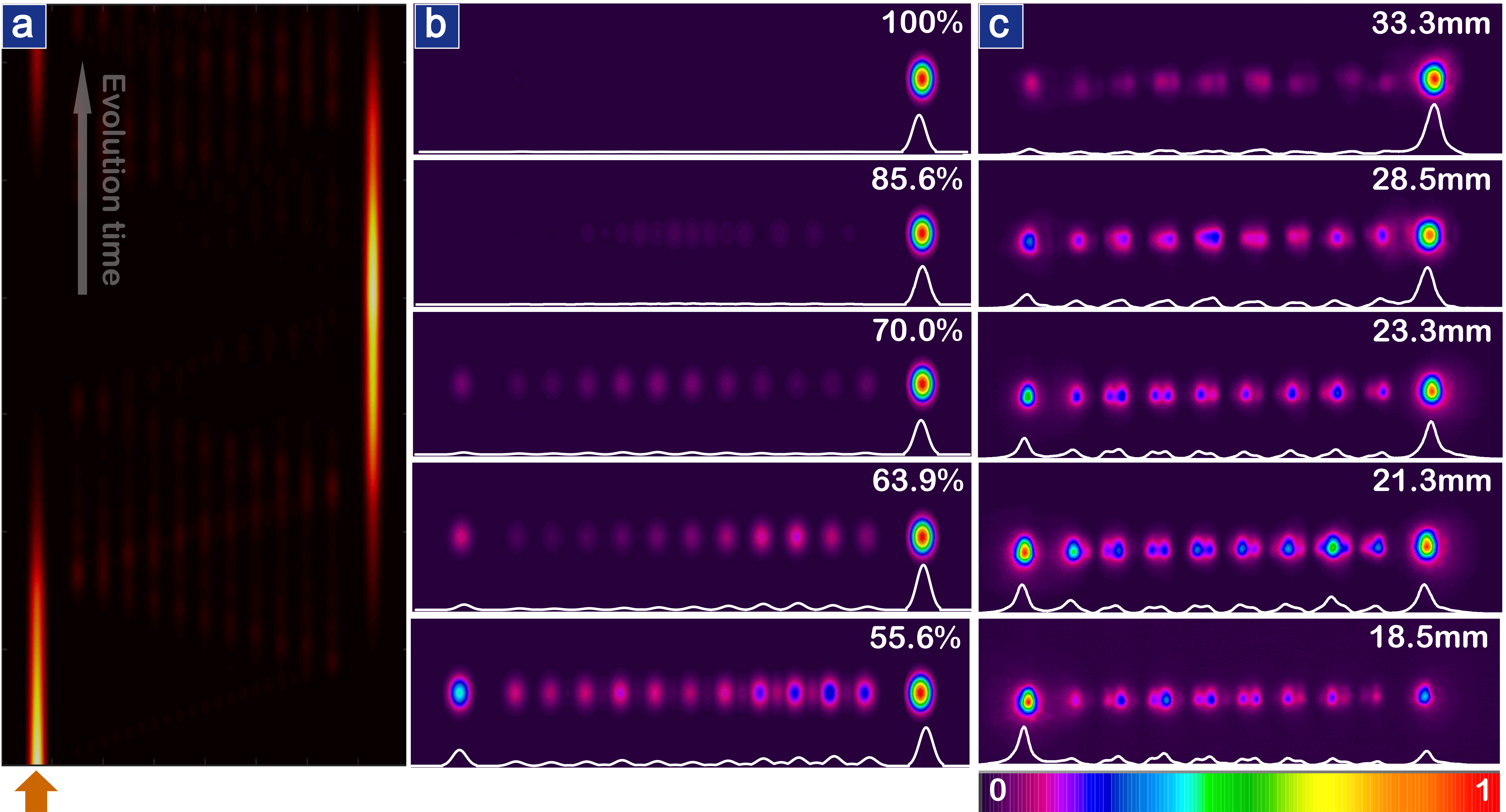}
\caption{\textbf{Theoretical and experimental results of time-evolution and transverse intensity distribution.} \textbf{a.} Numerical results of light propagation in the system with two boundary coupling sites and $21$ strongly coupling sites. \textbf{b.} and \textbf{c.}Theoretical and experimental results of the transverse intensity distribution in five different propagation lengths($18.5mm$, $21.3mm$, $23.3mm$, $28.5mm$ and $33.3mm$).}
\label{fig:figure3}
\end{figure*}

\subsection{Implementation of photonic Newton's cradle}
\vskip -3.5mm
Our aim of energy transport is to transfer photonic states, excited at the first site through the chain and retrieved at the remotely separated receiving site $N$ with a high efficiency $\eta$. Efficiency peaks when evolution time $t$ reaches the optimal $\tau$. In our experiment, we set the parameter $\alpha=0.12<1/\sqrt{23}$ and design the boundary controlled photonic lattices. According to the characterized coupling strength with the injection of coherent light at $810$nm (see Fig. \ref{fig:figure2}a and Appendix B), the strong coupling strength $J$ is chosen at $1.526/mm$ in a uniform site pitch of $5\mu m$ between $21$ nearest-neighbour sites. The boundary coupling strength is chosen at $0.189/mm$ in the site pitch of $16\mu m$ to the ends of the chain. 

The chain is able to transfer energy between two remote sites with the same fashion of energy exchange and interaction mechanics of Newton's cradle:
\begin{equation}
H=J_{eff} a_{1}^{\dagger} a_{N} + c.c.
\end{equation}
where effective coupling coefficient $J_{eff}$ is derived by $\alpha J/\sqrt{N-1}$ and is also shown in Fig. \ref{fig:figure2}a. Previous works deduce the practical meaning of the effective coupling coefficient $J_{eff}$ only from a theoretical view and here we perform a direct experimental observation. The effective coupling coefficient is only related to the weak coupling coefficient $\alpha J$, and can be obtained by measuring the optimal receiving time $\tau$ according to Eq. \ref{time}. As is shown in Fig. \ref{fig:figure2}b, we experimentally keep $J$ constant and tune $\alpha J$ by setting $5$ different weak coupling separations from $12\mu m$ to $15\mu m$ spacing $1\mu m$. With linear fitting we obtain a slope of $5.470$, which indicates that the experimentally obtained strong coupling strength $J$ is up to $1.316/mm$.

For a wide range that the boundary controlled condition is satisfied, the whole system behaves like a direct interaction between two remote sites, just like Newton's cradle, i.e. without considering the chain at middle. We fix the weak coupling coefficient $\alpha J$ and adjust finely the site pitch from $5\mu m$ to $5.6\mu m$ to acquire the strong coupling strength $J$ from $J|_{p=5}$ to $J|_{p=5.6}$. We measure the transverse intensity distribution while scanning the propagation length, which represents the evolution time $t$ due to the constant velocity on chip. As is shown in the shading period in Fig. \ref{fig:figure2}c, a trend fitting of experimental data gives a same $\tau$. Allowing for some marginal error, a trend fitting of experimental data gives a same $\tau$. Our measurement supports the theory that the optimal receiving time has little to do with the strong coupling strength and mainly relies on the weak coupling strength.



We visualize the time evolution of coherent light in the system by showing the imaged transverse intensity distribution at $5$ different transfer time (see Fig. \ref{fig:figure3}). Experimentally this is implemented by fabricating $5$ samples with different transfer time. The probability of the excitation at the $i_{th}$ site is given by $p_i(t)=|\braket{i}{\Phi(t)}|^2$, where $p_1(0)=1$. An optimal receiving time for faithful transfer can be found at the transfer time of $33.3mm$. The direct observed optimal time slightly deviates away from the value of $40.25mm$ predicted with the characterized coupling strength, which is due to the fact that the coupling strength does not rigorously follow the exponential relation to the propagation length in the very near coupling region. In addition, the next-nearest neighbor coupling, fabrication imperfection and even artificially introduced defects may also affect the real transfer performance.

\begin{figure*}
\includegraphics[width=0.7\textwidth]{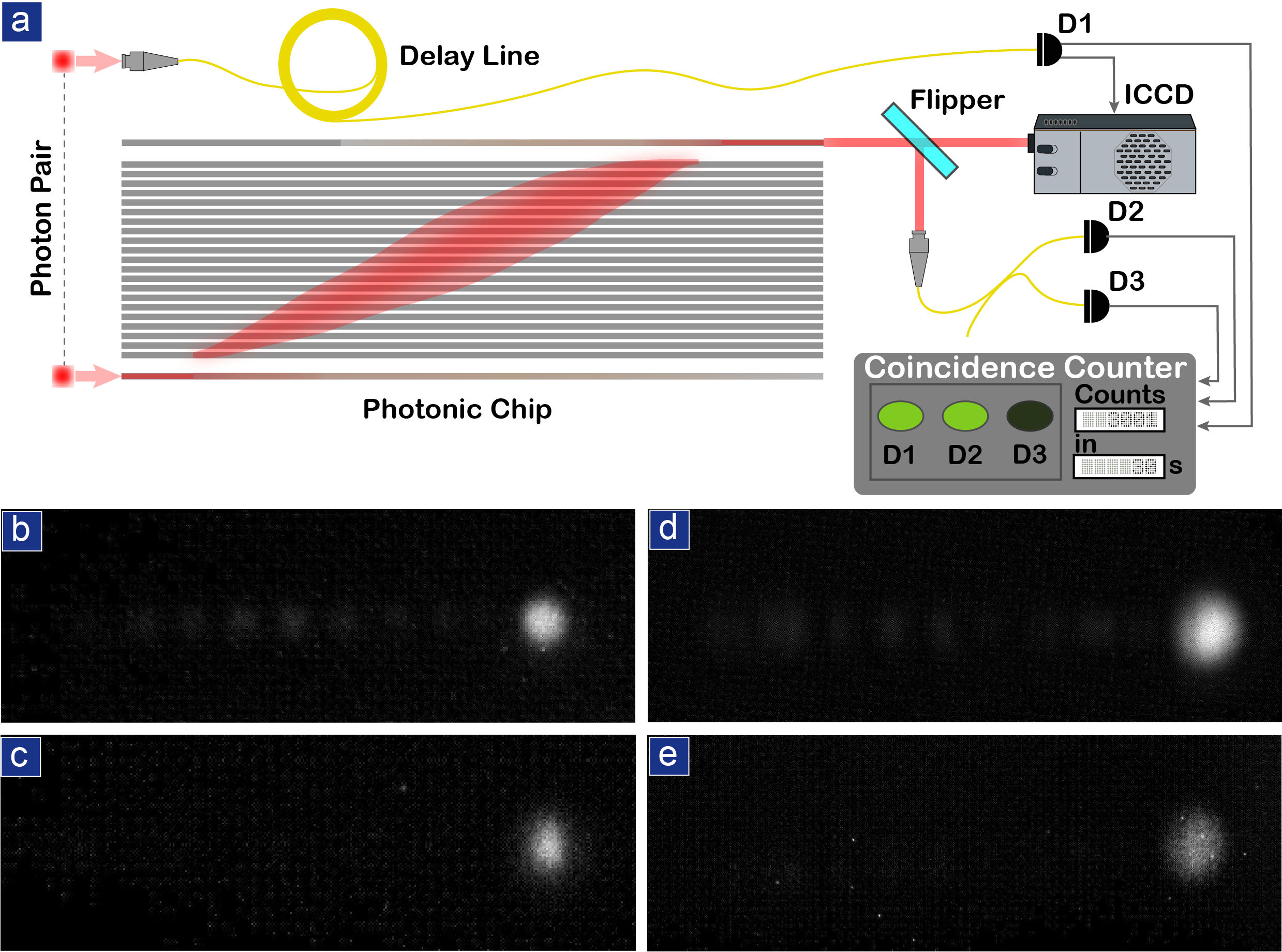}
\caption{\textbf{Single-photon Newton's cradle.} \textbf{a.} Experimental setup. A flipper mirror allows the detection system to be switched from direct single-photon imaging to Hanbury-Brown-Twiss interference. \textbf{b.} and \textbf{c.} Measured output intensity distribution of thermal light and heralded single photons for a standard boundary-controlled chain. \textbf{d.} and \textbf{e.} Measured output intensity distribution of thermal light and heralded single photons with an additional island site on the top middle of the chain.}
\label{fig:figure4}
\end{figure*}
\vskip -3.5mm
\subsection{Quantum correlation verification}
\vskip -3.5mm
Besides fundamental interest on the analog to Newton's cradle and the offered novel approach for energy transport, the demonstrated boundary-controlled chain also provides an elegant way to mediate long-range interactions on a photonic chip for quantum applications. To confirm its compatibility to quantum technologies, we inject heralded single photons instead of coherent light and measure the intensity distribution with single-photon imaging technique (see Fig. \ref{fig:figure4}a and APPENDIX C). The injected photon from single arm of our photon pair is actually a thermal light. With the trigger of a successful registration of the other photon, we are able to measure the output intensity distribution of genuine heralded single photons. For single-party injection, there is no difference on the optimal receiving time for coherent light, thermal light and single photons (see Fig. \ref{fig:figure4}b and \ref{fig:figure4}c).

In order to verify whether quantum correlation can be preserved in the evolution, we employ Hanbury-Brown-Twiss interferometer shown in Fig. \ref{fig:figure4}a to measure photon statistics of the output states. The anti-correlation function 
\begin{equation}
g^{(2)}(0)=\frac{p_1p_{123}}{p_{12}p_{13}}
\label{eq_g2_0}
\end{equation}
tends to $0$ for an ideally prepared single-photon state and tends to larger than 1 for classical light\cite{Spring2013}. Here, $p_1$ means the probability of a count in Detector $1$ and $p_{12}$, $p_{13}$ and $p_{123}$ represent the probabilities of simultaneous counts in those detectors. We observe an anti-correlation up to $0.0089\pm0.0019$, which suggests that single-photon property can be well preserved through the single-photon Newton's cradle.

\begin{figure*}
	\includegraphics[width=0.98\textwidth]{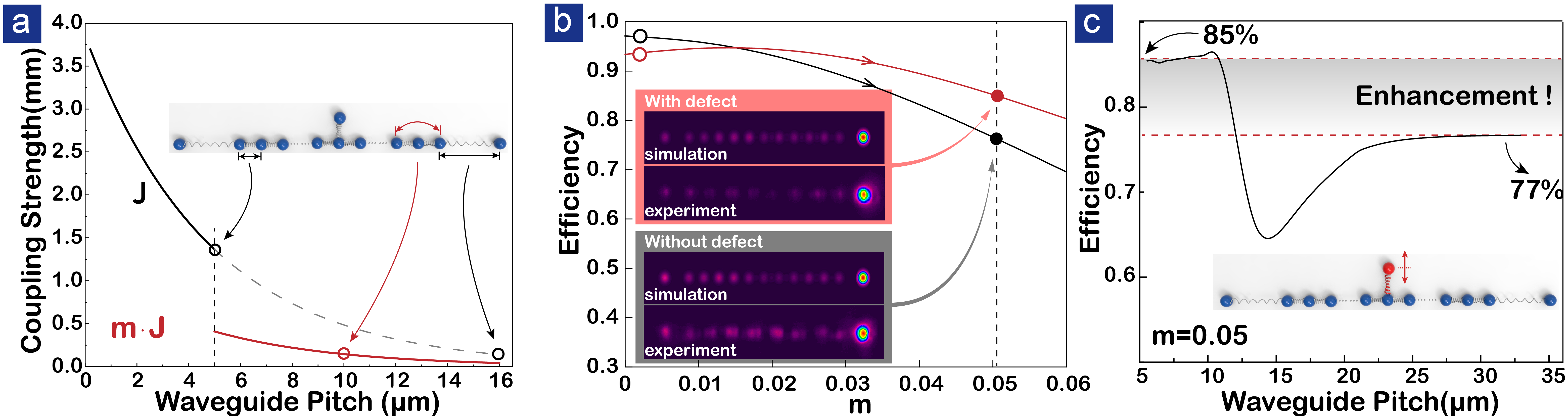}
	\caption{\textbf{Theoretical explanation of defect-induced efficiency enhancement.} \textbf{a.} The black line indicates the decay rate of coupling strength against pitch, and the red one shows the scenario where one waveguide is laid between them. $m$ is the attenuation ratio of the coupling strength when a waveguide is laid in the middle. \textbf{b.} Tuning $m$ gradually, we find a long period where a defect enhances the transport process and the inset shows the comparison between experimental results and their modified simulation counterparts. \textbf{c.} Taking $m$ as $0.05$, we simulate the transfer efficiency when the defect is moved close to and far away from the chain as is shown in the inset. We mark the experimental results obtained with and without a defect on the simulated curve, and find that they are well consistent with our theoretical modeling.
	}
	\label{fig:figure5}
\end{figure*}
\vskip -3.5mm
\subsection{Defect-induced enhancement of efficiency}
\vskip -3.5mm
Recently, the coherence and noise have been identified their existence and key roles in energy transport processes, for instance, in a photosynthesis complex \cite{Engel2007,Mohseni2018}, DNA \cite{Rieper2010}, and odor sensing \cite{Arndt2009} etc, in which we may have to explain the inner mechanics in quantum regime. Counterintuitively, early work suggested that the noise and decoherence may have a positive influence on quantum processes \cite{Plenio2002,Caruso2016,Mohseni2018,Maziero2009,Oh2012,Stolze2014}. Meanwhile, theoretical investigation has also shown that the boundary controlled chain may be a fault-tolerant system \cite{Wang2013}. We construct a nonideal scenario by adding an island site next to the central site with the same separation as other strong coupled sites. The newly added island site can be considered as noise, defect or environment for the original boundary-controlled chain. The imaged output intensity distributions with thermal light and heralded single photons are shown in Fig. \ref{fig:figure4}d and Fig. \ref{fig:figure4}e, respectively. We do observe an enhancement of $8\%$ with an efficiency up to $86.20\%$, though the maximal efficiency is obtained at a bit longer optimal receiving time of $35.4mm$. Hanbury-Brown-Twiss experiment also suggest a good preservation of quantum correlation with an measured anti-correlation up to $0.027\pm0.0032$.

In a general case, the nearest coupling is only considered \cite{Stolze2014} but in an experimental environment, our linear arrangement of waveguides have a very near pitch of $5 \mu m$ between each other, which means the next-nearest pitch is $10\mu m$ and it is also non-negligible. Furthermore, we will show that one waveguide laid in the middle will largely affects the propagation of evanescent wave. We therefore construct a modified model to explain the observed defect-induced enhancement of efficiency.

We consider that the coupling strength attenuates to $m$ times ($0<m<1$) when a waveguide is laid in the middle. In consideration of the same propagation media, for the coupling coefficient $J'$ plotted in red in Fig. \ref{fig:figure5}a, the exponential decay constant $k$ of coupling strength against waveguide pitch $p$ remains the same with that for $J$ ploted in black in Fig. \ref{fig:figure5}a, i.e. $J = e^{-k p}$ and $J' = m e^{-k p}$. The modified Hamiltonian $H'$ is expanded as follows: 
\begin{equation}
H'=H + (\sum_{i=2}^{N-3}m J a_{i}^{\dagger} a_{i+2} + c.c.)
\end{equation}
where $H$ is the Hamiltonian in Eq. \ref{eq:Equation1}. We can find a continuously differential transfer efficiency (Fig. \ref{fig:figure5}b) when the parameter $m$ is adjusted dynamically from $0$ to $0.06$. With the $m$ value of $0.05$, we can obtain an increase of $8\%$ from $77\%$ (without defect) to $85\%$ (with defect).


We may find that the defect can play a dual role in the model of photonic Newton's cradle. Ideally, the transfer efficiency should reach a unity. In practice, however, the defects introduced by fabrication or environment can induce a drop away from the expected value. Interestingly, we may also improve the transfer efficiency by the defect itself if we can artificially introduce the defect appropriate in light of a standard coupling mode theory.
\vskip -3.5mm
\section{CONCLUSION}
\vskip -3.5mm
In summary, we present an experimental demonstration of Newton's cradle by using a boundary-controlled chain on a photonic chip. Energy transport can be directly conducted between two remote sites with the same fashion of energy exchange and interaction mechanics in Newton's cradle. A long-range interaction is mediated by a long chain composed of $21$ strongly coupled sites and a high retrieval efficiency can be obtained in both uniform and defect-doped chain structures. Our results may offer a new approach of flexible Hamiltonian engineering beyond geometric limitation, such as to bypass faulty nodes or to bridge remote terminals, enabling on-demand design and construction integrated quantum networks for quantum simulation.
\vskip -3.5mm
\vskip -3.5mm
\section*{ACKNOWLEDGMENTS}
\vskip -3.5mm
The authors thank Roberto Osellame, Chang-Pu Sun and Jian-Wei Pan for helpful discussions. This work was supported by National Key R\&D Program of China (2017YFA0303700); National Natural Science Foundation of China (NSFC) (61734005, 11761141014, 11690033); Science and Technology Commission of Shanghai Municipality (STCSM) (15QA1402200, 16JC1400405, 17JC1400403); Shanghai Municipal Education Commission (SMEC)(16SG09, 2017-01-07-00-02-E00049); X.-M.J. acknowledges support from the National Young 1000 Talents Plan. L.A.W. is supported by the Basque Government (grant IT472-10) and the Spanish MICINN (Project No. FIS2012-36673- C03-03).

\vskip -3.5mm
\vskip -3.5mm
\section*{APPENDIX A: FABRICATION DETAILS}
\label{appendix:A}
\vskip -3.5mm
An ultrafast pulse focused into transparent materials can produce a permanent refractive index increase due to the localized nonlinear absorption process. A waveguide can be constructed with a translation of photonic wafer in three dimensions. The femtosecond laser ($10$W, $1026$nm) with $290$fs pulse duration and $1$MHz repetition rate is frequency doubled to 513nm, and is further fed into an spatial light modulator to create burst trains. By using a $50\times$ objective lens with a numerical aperture of $0.55$, we focus the laser on a borosilicate substrate ($100mm \times 20mm \times 1mm$) to fabricate the chain structure consisting of $23$ waveguides, each of which has an elliptical cross section about $7\mu m \times 5\mu m$.
\vskip -3.5mm
\vskip -3.5mm
\section*{APPENDIX B: COUPLING STRENGTH}
\label{appendix:B}
\vskip -3.5mm
The coupling strength between neighboring sites exponentially decreases with the site pitch as $J=3.944\times e^{-0.1899\times p}$, where $p$ is the site pitch in the unit of $\mu m$. The strong coupling strength in the site train is about $1.526/mm$ in the uniform site pitch of $5\mu m$ among $21$ nearest neighbour sites. The boundary coupling strength is about $0.1890/mm$ in the site pitch of $16\mu m$ between the sending (or receiving) site and the nearest neighboring site.
\vskip -3.5mm
\vskip -3.5mm
\section*{APPENDIX C: DETAILS OF SINGLE-PHOTON EXPERIMENT}
\label{appendix:C}
\vskip -3.5mm
A single photon at $810 nm$, heralded from a spontaneous parametric down conversion photon pair, is injected into the sending site in the photonic chip instead of classical light. Intensified charge coupled device (ICCD) cameras with a gate width of $4.5ns$ has a very good signal-to-noise ratio and single-photon sensitivity. The single-photon events can be captured at the output end of the photonic chip (shown in Fig. \ref{fig:figure4}a). The other photon is delayed by an coiled optical fiber before registering at an avalanche photodiode D1. With the trigger from D1, we accumulate $3000$ frames in $30$ minutes to obtain the heralded single-photon results (shown in Fig. \ref{fig:figure4}c). Without the trigger, one arm of photon pair is actually in a single-photon-level thermal state. The measured output intensity distribution is shown in Fig. \ref{fig:figure4}b.

We use a Hanbury-Brown-Twiss scheme to verify the ability of preserving the properties of single photons. By turning up the flipper mirror, photons can be collected into a fiber beam splitter. The two avalanche photodiodes D1 and D2 count photons at the two outputs of beam splitter and give the probability $p_2$ and $p_3$. Together with the probability $p_1$ obtained from D1, we can evaluate the intensity cross-correlation function by
\begin{equation}
g_{si}=\frac{(p_{12}+p_{13})}{p_{1}(p_{2}+p_{3})}
\end{equation}
Besides, the anti-correlation $g^{(2)}(0)$ of heralded single photons ideally tends to zero and the mathematical expression is obtained in Eq. \ref{eq_g2_0} 

The cross-correlation and anti-correlation are found to reach $g_{si}=148.99\pm1.96$ and $g_H=0.0089\pm0.0019$ respectively, which suggest a good preservation of quantum correlation.


\begin{thebibliography}{10}
	\expandafter\ifx\csname url\endcsname\relax
	\def\url#1{\texttt{#1}}\fi
	\expandafter\ifx\csname urlprefix\endcsname\relax\def\urlprefix{URL }\fi
	\providecommand{\bibinfo}[2]{#2}
	\providecommand{\eprint}[2][]{\url{#2}}
	
	\bibitem{Engel2007}
	\bibinfo{author}{Engel, G.~S.} \emph{et~al.}
	\newblock \bibinfo{title}{Evidence for wavelike energy transfer through quantum coherence in photosynthetic systems}.
	\newblock \emph{\bibinfo{journal}{Nature}} \textbf{\bibinfo{volume}{446}},
	\bibinfo{pages}{782} (\bibinfo{year}{2007}).
	
	\bibitem{Mohseni2018}
	\bibinfo{author}{Mohseni, M.}, \bibinfo{author}{Rebentrost, P.},
	\bibinfo{author}{Lloyd, S.} \& \bibinfo{author}{Aspuru-Guzik, A.}
	\newblock \bibinfo{title}{Environment-assisted quantum walks in photosynthetic energy transfer}.
	\newblock \emph{\bibinfo{journal}{J. Chem. Phys.}}
	\textbf{\bibinfo{volume}{129}}, \bibinfo{pages}{174106}
	(\bibinfo{year}{2018}).
	
	\bibitem{Rieper2010}
	\bibinfo{author}{Rieper, E.}, \bibinfo{author}{Anders, J.} \&
	\bibinfo{author}{Vedral, V.}
	\newblock \bibinfo{title}{Quantum entanglement between the electron clouds of nucleic acids in dna}.
	\newblock \emph{\bibinfo{journal}{arXiv: 1006.4053}}
	(\bibinfo{year}{2010}).
	
	\bibitem{Arndt2009}
	\bibinfo{author}{Arndt, M.}, \bibinfo{author}{Juffmann, T.} \&
	\bibinfo{author}{Vedral, V.}
	\newblock \bibinfo{title}{Quantum physics meets biology}.
	\newblock \emph{\bibinfo{journal}{HFSP Journal}} \textbf{\bibinfo{volume}{3}},
	\bibinfo{pages}{386--400} (\bibinfo{year}{2009}).
	
	\bibitem{Bentivegna2015}
	\bibinfo{author}{Bentivegna, M.} \emph{et~al.}
	\newblock \bibinfo{title}{Experimental scattershot boson sampling}.
	\newblock \emph{\bibinfo{journal}{Sci. Adv.}} \textbf{\bibinfo{volume}{1}},
	\bibinfo{pages}{e1400255} (\bibinfo{year}{2015}).
	
	\bibitem{Pitsios2017}
	\bibinfo{author}{Pitsios, I.} \emph{et~al.}
	\newblock \bibinfo{title}{Photonic simulation of entanglement growth and
		engineering after a spin chain quench}.
	\newblock \emph{\bibinfo{journal}{Nat. Commun.}}
	\textbf{\bibinfo{volume}{8}}, \bibinfo{pages}{1569} (\bibinfo{year}{2017}).
	
	\bibitem{Kempe2003}
	\bibinfo{author}{Kempe, J.}
	\newblock \bibinfo{title}{Quantum random walks: An introductory overview}.
	\newblock \emph{\bibinfo{journal}{Contemp. Phys.}}
	\textbf{\bibinfo{volume}{44}}, \bibinfo{pages}{307--327}
	(\bibinfo{year}{2003}).
	
	\bibitem{Schwartz2007}
	\bibinfo{author}{Schwartz, T.}, \bibinfo{author}{Bartal, G.},
	\bibinfo{author}{Fishman, S.} \& \bibinfo{author}{Segev, M.}
	\newblock \bibinfo{title}{Transport and anderson localization in disordered
		two-dimensional photonic lattices}.
	\newblock \emph{\bibinfo{journal}{Nature}} \textbf{\bibinfo{volume}{446}},
	\bibinfo{pages}{52} (\bibinfo{year}{2007}).
	
	\bibitem{Lahini2008}
	\bibinfo{author}{Lahini, Y.} \emph{et~al.}
	\newblock \bibinfo{title}{Anderson localization and nonlinearity in
		one-dimensional disordered photonic lattices}.
	\newblock \emph{\bibinfo{journal}{Phys. Rev. Lett.}}
	\textbf{\bibinfo{volume}{100}}, \bibinfo{pages}{013906}
	(\bibinfo{year}{2008}).
	
	\bibitem{Maziero2009}
	\bibinfo{author}{Maziero, J.}, \bibinfo{author}{Céleri, L.~C.},
	\bibinfo{author}{Serra, R.~M.} \& \bibinfo{author}{Vedral, V.}
	\newblock \bibinfo{title}{Classical and quantum correlations under
		decoherence}.
	\newblock \emph{\bibinfo{journal}{Phys. Rev. A}} \textbf{\bibinfo{volume}{80}},
	\bibinfo{pages}{044102} (\bibinfo{year}{2009}).
	
	\bibitem{Christandl2004}
	\bibinfo{author}{Christandl, M.}, \bibinfo{author}{Datta, N.},
	\bibinfo{author}{Ekert, A.} \& \bibinfo{author}{Landahl, A.~J.}
	\newblock \bibinfo{title}{Perfect state transfer in quantum spin networks}.
	\newblock \emph{\bibinfo{journal}{Phys. Rev. Lett.}}
	\textbf{\bibinfo{volume}{92}}, \bibinfo{pages}{187902}
	(\bibinfo{year}{2004}).
	
	\bibitem{Bellec2012}
	\bibinfo{author}{Bellec, M.}, \bibinfo{author}{Nikolopoulos, G.~M.} \&
	\bibinfo{author}{Tzortzakis, S.}
	\newblock \bibinfo{title}{Faithful communication hamiltonian in photonic
		lattices}.
	\newblock \emph{\bibinfo{journal}{Opt. Lett.}} \textbf{\bibinfo{volume}{37}},
	\bibinfo{pages}{4504--4506} (\bibinfo{year}{2012}).
	
	\bibitem{Chapman2016}
	\bibinfo{author}{Chapman, R.~J.} \emph{et~al.}
	\newblock \bibinfo{title}{Experimental perfect state transfer of an entangled
		photonic qubit}.
	\newblock \emph{\bibinfo{journal}{Nat. Commun.}} \textbf{\bibinfo{volume}{7}},
	\bibinfo{pages}{11339} (\bibinfo{year}{2016}).
	
	\bibitem{Wojcik2005}
	\bibinfo{author}{Wójcik, A.} \emph{et~al.}
	\newblock \bibinfo{title}{Unmodulated spin chains as universal quantum wires}.
	\newblock \emph{\bibinfo{journal}{Phys. Rev. A}} \textbf{\bibinfo{volume}{72}},
	\bibinfo{pages}{034303} (\bibinfo{year}{2005}).
	
	\bibitem{Yao2011}
	\bibinfo{author}{Yao, N.~Y.} \emph{et~al.}
	\newblock \bibinfo{title}{Robust quantum state transfer in random unpolarized
		spin chains}.
	\newblock \emph{\bibinfo{journal}{Phys. Rev. Lett.}}
	\textbf{\bibinfo{volume}{106}}, \bibinfo{pages}{040505}
	(\bibinfo{year}{2011}).
	
	\bibitem{Wang2013}
	\bibinfo{author}{Wang, Z.-M.}, \bibinfo{author}{Wu, L.-A.},
	\bibinfo{author}{Modugno, M.}, \bibinfo{author}{Yao, W.} \&
	\bibinfo{author}{Shao, B.}
	\newblock \bibinfo{title}{Fault-tolerant almost exact state transmission}.
	\newblock \emph{\bibinfo{journal}{Sci. Rep.}} \textbf{\bibinfo{volume}{3}},
	\bibinfo{pages}{3128} (\bibinfo{year}{2013}).
	
	\bibitem{Wu2002}
	\bibinfo{author}{Wu, L.-A.} \& \bibinfo{author}{Lidar, D.~A.}
	\newblock \bibinfo{title}{Power of anisotropic exchange interactions:
		Universality and efficient codes for quantum computing}.
	\newblock \emph{\bibinfo{journal}{Phys. Rev. A}} \textbf{\bibinfo{volume}{65}},
	\bibinfo{pages}{042318} (\bibinfo{year}{2002}).
	
	\bibitem{Plenio2005}
	\bibinfo{author}{Plenio, M.~B.} \& \bibinfo{author}{Semião, F.~L.}
	\newblock \bibinfo{title}{High efficiency transfer of quantum information and
		multiparticle entanglement generation in translation-invariant quantum
		chains}.
	\newblock \emph{\bibinfo{journal}{New J. Phys.}}
	\textbf{\bibinfo{volume}{7}}, \bibinfo{pages}{73} (\bibinfo{year}{2005}).
	
	\bibitem{Shi2005}
	\bibinfo{author}{Shi, T.}, \bibinfo{author}{Li, Y.}, \bibinfo{author}{Song, Z.}
	\& \bibinfo{author}{Sun, C.-P.}
	\newblock \bibinfo{title}{Quantum-state transfer via the ferromagnetic chain in
		a spatially modulated field}.
	\newblock \emph{\bibinfo{journal}{Phys. Rev. A}} \textbf{\bibinfo{volume}{71}},
	\bibinfo{pages}{032309} (\bibinfo{year}{2005}).
	
	\bibitem{Li2005}
	\bibinfo{author}{Li, Y.}, \bibinfo{author}{Shi, T.}, \bibinfo{author}{Chen,
		B.}, \bibinfo{author}{Song, Z.} \& \bibinfo{author}{Sun, C.-P.}
	\newblock \bibinfo{title}{Quantum-state transmission via a spin ladder as a
		robust data bus}.
	\newblock \emph{\bibinfo{journal}{Phys. Rev. A}} \textbf{\bibinfo{volume}{71}},
	\bibinfo{pages}{022301} (\bibinfo{year}{2005}).
	
	\bibitem{CamposVenuti2007}
	\bibinfo{author}{Campos~Venuti, L.}, \bibinfo{author}{Degli Esposti~Boschi, C.}
	\& \bibinfo{author}{Roncaglia, M.}
	\newblock \bibinfo{title}{Qubit teleportation and transfer across
		antiferromagnetic spin chains}.
	\newblock \emph{\bibinfo{journal}{Phys. Rev. Lett.}}
	\textbf{\bibinfo{volume}{99}}, \bibinfo{pages}{060401}
	(\bibinfo{year}{2007}).
	
	\bibitem{Oh2012}
	\bibinfo{author}{Oh, S.}, \bibinfo{author}{Shim, Y.-P.}, \bibinfo{author}{Fei,
		J.}, \bibinfo{author}{Friesen, M.} \& \bibinfo{author}{Hu, X.}
	\newblock \bibinfo{title}{Effect of randomness on quantum data buses of
		heisenberg spin chains}.
	\newblock \emph{\bibinfo{journal}{Phys. Rev. B}} \textbf{\bibinfo{volume}{85}},
	\bibinfo{pages}{224418} (\bibinfo{year}{2012}).
	
	\bibitem{Stolze2014}
	\bibinfo{author}{Stolze, J.}, \bibinfo{author}{Álvarez, G.~A.},
	\bibinfo{author}{Osenda, O.} \& \bibinfo{author}{Zwick, A.}
	\newblock \bibinfo{title}{Robustness of spin-chain state-transfer schemes}.
	\newblock In \bibinfo{editor}{Nikolopoulos, G.~M.} \& \bibinfo{editor}{Jex, I.}
	(eds.) \emph{\bibinfo{booktitle}{Quantum State Transfer and Network
			Engineering}}, \bibinfo{pages}{149--182} (\bibinfo{publisher}{Springer Berlin
		Heidelberg}, \bibinfo{address}{Berlin, Heidelberg}, \bibinfo{year}{2014}).
	
	\bibitem{Stolze2016}
	\bibinfo{author}{Stolze, J.} \& \bibinfo{author}{Zenchuk, A.~I.}
	\newblock \bibinfo{title}{Remote two-qubit state creation and its robustness}.
	\newblock \emph{\bibinfo{journal}{Quant. Inf. Process}}
	\textbf{\bibinfo{volume}{15}}, \bibinfo{pages}{3347--3366}
	(\bibinfo{year}{2016}).
	
		\bibitem{Vieira2018}
	\bibinfo{author}{Vieira, R.} \& \bibinfo{author}{Rigolin, G.}
	\newblock \bibinfo{title}{Almost perfect transport of an entangled two-qubit
		state through a spin chain}.
	\newblock \emph{\bibinfo{journal}{Phys. Lett. A}}
	\textbf{\bibinfo{volume}{382}}, \bibinfo{pages}{2586--2594}
	(\bibinfo{year}{2018}).

	\bibitem{Szameit2007}
	\bibinfo{author}{Szameit, A.}, \bibinfo{author}{Dreisow, F.},
	\bibinfo{author}{Pertsch, T.}, \bibinfo{author}{Nolte, S.} \&
	\bibinfo{author}{Tünnermann, A.}
	\newblock \bibinfo{title}{Control of directional evanescent coupling in fs
		laser written waveguides}.
	\newblock \emph{\bibinfo{journal}{Opt. Express}} \textbf{\bibinfo{volume}{15}},
	\bibinfo{pages}{1579--1587} (\bibinfo{year}{2007}).
	
	\bibitem{Feng2016}
	\bibinfo{author}{Feng, Z.} \emph{et~al.}
	\newblock \bibinfo{title}{Invisibility cloak printed on a photonic chip}.
	\newblock \emph{\bibinfo{journal}{Sci. Rep.}}
	\textbf{\bibinfo{volume}{6}}, \bibinfo{pages}{28527} (\bibinfo{year}{2016}).
	
	\bibitem{Tang2018}
	\bibinfo{author}{Tang, H.} \emph{et~al.}
	\newblock \bibinfo{title}{Experimental quantum fast hitting on hexagonal
		graphs}.
	\newblock \emph{\bibinfo{journal}{Nat. Photon.}}
	\textbf{\bibinfo{volume}{12}}, \bibinfo{pages}{754--758}
	(\bibinfo{year}{2018}).
	
	\bibitem{Tang2018a}
	\bibinfo{author}{Tang, H.} \emph{et~al.}
	\newblock \bibinfo{title}{Experimental two-dimensional quantum walk on a
		photonic chip}.
	\newblock \emph{\bibinfo{journal}{Sci. Adv.}} \textbf{\bibinfo{volume}{4}},
	\bibinfo{pages}{eaat3174} (\bibinfo{year}{2018}).
	
	\bibitem{Nielsen2002}
	\bibinfo{author}{Nielsen, M.~A.} \& \bibinfo{author}{Chuang, I.}
	\newblock \bibinfo{title}{Quantum computation and quantum information}
	(\bibinfo{year}{2002}).
	
	\bibitem{Zwick2012}
	\bibinfo{author}{Zwick, A.}, \bibinfo{author}{Álvarez, G.~A.},
	\bibinfo{author}{Stolze, J.} \& \bibinfo{author}{Osenda, O.}
	\newblock \bibinfo{title}{Spin chains for robust state transfer: Modified
		boundary couplings versus completely engineered chains}.
	\newblock \emph{\bibinfo{journal}{Phys. Rev. A}} \textbf{\bibinfo{volume}{85}},
	\bibinfo{pages}{012318} (\bibinfo{year}{2012}).
	
	\bibitem{Spring2013}
	\bibinfo{author}{Spring, J.~B.} \emph{et~al.}
	\newblock \bibinfo{title}{On-chip low loss heralded source of pure single
		photons}.
	\newblock \emph{\bibinfo{journal}{Opt. Express}} \textbf{\bibinfo{volume}{21}},
	\bibinfo{pages}{13522--13532} (\bibinfo{year}{2013}).
		
	\bibitem{Plenio2002}
	\bibinfo{author}{Plenio, M.~B.} \& \bibinfo{author}{Huelga, S.~F.}
	\newblock \bibinfo{title}{Entangled light from white noise}.
	\newblock \emph{\bibinfo{journal}{Phys. Rev. Lett.}}
	\textbf{\bibinfo{volume}{88}}, \bibinfo{pages}{197901}
	(\bibinfo{year}{2002}).
	
	\bibitem{Caruso2016}
	\bibinfo{author}{Caruso, F.}, \bibinfo{author}{Crespi, A.},
	\bibinfo{author}{Ciriolo, A.~G.}, \bibinfo{author}{Sciarrino, F.} \&
	\bibinfo{author}{Osellame, R.}
	\newblock \bibinfo{title}{Fast escape of a quantum walker from an integrated
		photonic maze}.
	\newblock \emph{\bibinfo{journal}{Nat. Commun.}} \textbf{\bibinfo{volume}{7}},
	\bibinfo{pages}{11682} (\bibinfo{year}{2016}).
	
\end{thebibliography}

\end{document}